\documentclass{article}

\usepackage{fancyvrb,fvextra}
\usepackage{xcolor} 

\usepackage{microtype}
\usepackage{graphicx}
\usepackage{subfigure}
\usepackage{booktabs} 
\usepackage{color}
\usepackage{algpseudocode}
\usepackage{algorithm}

\usepackage{natbib} 
\usepackage{hyperref}
\usepackage{ulem}
\usepackage{amsmath}
\usepackage{amssymb}  
\usepackage{mathtools}
\usepackage{amsthm}
\usepackage{cleveref}
\usepackage{bm}
\usepackage{tcolorbox}
\usepackage{listings}

\hypersetup{
  colorlinks=true,
  breaklinks=true,
  linkcolor=blue,
  citecolor=blue,
  urlcolor=blue,
  pdftitle={Agentic Inequality},
  pdfauthor={Matthew Sharp, Omer Bilgin, Iason Gabriel, and Lewis Hammond},
  pdfkeywords={AI agents, inequality, governance, agentic inequality}
}

\DefineVerbatimEnvironment{MyVerbatim}{Verbatim}{
  commandchars=@\{\},
  breaklines=true
}

\lstdefinestyle{customstyle}{
    moredelim={[is][keywordstyle]{@@}{@@}},  
    keywordstyle=\color{blue}\textbf,               
    breaklines=true,  
    basicstyle=\ttfamily
}

\newtcolorbox{mybox}[1][]{
    title=#1,
    fonttitle=\small,
    fontupper=\small,
    left=2mm,
    right=2mm,
    top=1mm,
    bottom=0mm,
}

\crefname{observation}{Observation}{Observations}

\def\1{\mathbf{1}}

\usepackage{fullpage}

\title{The Compulsory Imaginary: AGI and Corporate Authority}

\author{Emilio Barkett\thanks{Columbia University. Correspondence: \texttt{eab2291@columbia.edu}}}

\date{} 

\begin{document}

\maketitle

\begingroup

\renewcommand{\thefootnote}{\arabic{footnote}}
\setcounter{footnote}{1} 

\endgroup

\begin{abstract}
The development of artificial general intelligence (AGI) is as much an imaginative project as a technical one. This paper argues that the two leading AGI firms---OpenAI and Anthropic---construct sociotechnical imaginaries through a structurally consistent rhetorical strategy, despite meaningful differences in execution. Drawing on \citet{Jasanoff-2015-Dreamscapes}'s framework of sociotechnical imaginaries, the sociology of expectations \citep{Borup-2006-sociology}, and the concept of the vanguard vision \citep{hilgartner-2015-science}, the paper analyzes two essays published within weeks of each other in late 2024: Sam Altman's \textit{The Intelligence Age} and Dario Amodei's \textit{Machines of Loving Grace}. A close comparative reading identifies four rhetorical operations shared across both texts: the self-exemption move, which disavows prophetic authority while exercising it; teleological naturalization, which embeds AGI's arrival in narratives of historical and moral inevitability; qualified acknowledgment, which absorbs genuine concessions to risk into an overarching optimistic frame; and implicit indispensability, which positions each firm as central to the imagined future without naming it as a commercial actor. That two competing institutions with different cultures, different risk philosophies, and leaders with notably different public personae nevertheless converge on the same underlying rhetorical architecture---despite strong competitive incentives to differentiate their public visions---suggests the imaginary is not only a firm-level communication strategy but also a feature of the institutional position these firms occupy. The paper makes three contributions: it extends the sociotechnical imaginaries framework from nation-states to private firms at the frontier of transformative technology development; it identifies the specific discursive mechanism through which corporate authority over technological futures is projected and oriented toward stabilization; and it demonstrates through comparative analysis that this mechanism is at minimum structural rather than purely idiosyncratic. The findings raise the question of what institutional arrangements would be required to make that authority contestable from outside the firms that produce it.
\end{abstract}





\section{Introduction}
\label{sec:intro}

In October 2024, Sam \citeauthor{altman-2024-intelligence}, the CEO of OpenAI, published a short essay predicting that humanity stood at ``the doorstep of the next leap in prosperity.'' The following month, Dario \citet{amodei-2024-machines}, the CEO of Anthropic, published a far longer essay imagining the defeat of most diseases, the doubling of the human lifespan, and what he called ``eternal 1991''---a world in which liberal democracy has permanently prevailed over authoritarianism. Neither essay announced a product release or disclosed a technical finding, but both were addressed to ``humanity,'' to ``our children,'' to ``the future.'' Both appeared on the personal websites of their authors and circulated widely across technical and popular media.

It is not incidental that two busy CEOs each chose to write public essays about the future of civilization. It is evidence of a constitutive feature of contemporary AGI development: building the technology and narrating its future are not separable activities. AI firms are not merely constructing systems, rather, they are constructing the social understanding of what those systems mean, what futures they make possible, and who is qualified to lead humanity toward them. The essays analyzed here are not supplements to the technical work, they are part of it.

I argue that OpenAI and Anthropic construct \textit{sociotechnical imaginaries} of AGI through a structurally consistent rhetorical strategy, despite meaningful differences in execution. Altman's essay performs the imaginary through prophetic brevity; Amodei's performs it through the simulation of scientific rigor. Yet both arrive at the same ideological destination, that AGI is historically inevitable, its arrival will be civilizationally transformative, and the firms developing it are indispensable to ensuring that transformation goes well. The structural convergence of this mechanism across two competing institutions (with different cultures, different risk philosophies, and leaders with notably different public personae) is the paper's central finding, and it suggests the imaginary is not a strategic communication choice but a feature of the institutional position these firms occupy.

The concept of the sociotechnical imaginary, developed by \citet{Jasanoff-2015-Dreamscapes, Jasanoff-2009-Containing}, describes collectively held, institutionally stabilized, and publicly performed visions of desirable futures animated by shared understandings of forms of social life attainable through advances in science and technology. Originally developed through comparative analysis of national techno-scientific cultures, the framework has since been extended to corporate actors \citep{Jasanoff-2015-Dreamscapes} and to public perceptions of AI \citep{sartori-2023-minding}. It is well suited to the present analysis because AI firms now occupy a position analogous to the state actors Jasanoff originally studied as they are the primary institutions with both the technical capacity and the resources to define what AGI is and what futures it enables. Yet unlike state actors, they bear no formal democratic accountability for the imaginaries they produce, a gap that is the paper's normative concern as much as its analytical one.

Existing STS scholarship on AI and public discourse has examined how non-expert publics receive and interpret AI narratives \citep{sartori-2023-minding}, how expectations structure technological development \citep{Borup-2006-sociology}, and how corporate actors use narrative to constitute stakeholders and markets \citep{van-1993-promising}. Less attention has been paid to the upstream question of how the authoritative narratives circulating in public discourse are produced in the first place. This paper takes up that production-side question, examining the specific discursive operations through which two leading firms construct and stabilize their imaginaries. The comparative approach is central to the argument. By analyzing both essays together, the paper distinguishes what is firm-specific from what is structural, and makes the stronger claim that the imaginary reflects position rather than personality.


\section{Theoretical Framework}
\label{sec:theory}

\subsection{Sociotechnical Imaginaries}

Sociotechnical imaginaries are ``collectively held, institutionally stabilized, and publicly performed visions of desirable futures, animated by shared understandings of forms of social life and social order attainable through, and supportive of, advances in science and technology'' \citep{Jasanoff-2009-Containing}. Three features of this definition are load-bearing for the present analysis.

First, imaginaries are normative rather than merely predictive. They encode visions of how life ought to be lived and what social arrangements are desirable, and in doing so they naturalize a particular distribution of authority by embedding it in a vision of the future that appears not only attainable but morally necessary \citep{Jasanoff-2015-Dreamscapes}. When Altman writes that ``everyone's life can be better than anyone's life is now'' \citep{altman-2024-intelligence}, the statement is not a prediction but a moral warrant, one that positions whoever is building that future as acting on behalf of humanity rather than on behalf of a firm.

Second, imaginaries are performative in the sense that \citeauthor{Jasanoff-2015-Dreamscapes} intend. They do not exist as privately held views but as social facts produced through sustained institutional enactment. A CEO essay is a paradigmatic performance of this kind. It is authoritative by virtue of who delivers it, public by virtue of where it appears, and institutionally backed by the firm whose leader speaks. The same claim made by an anonymous blogger carries no imaginary weight; made by the CEO of the firm building the system described, it is recruited into a collectively held vision. The performance is not incidental to the imaginary's content but constitutive of it.

Third, imaginaries naturalize contingency. Once stabilized, an imaginary ``makes its constitutive assumptions invisible'' \citep[p.~4]{Jasanoff-2015-Dreamscapes}, the vision of the future no longer appears as one possible outcome among many but as the expression of an underlying necessity. This naturalization is not achieved through explicit argument but through narrative form, the positioning of specific actors as history's designated agents, and the systematic exclusion of alternative trajectories from what is imaginable at all.

\subsection{The Sociology of Expectations and the Vanguard Vision}

The sociotechnical imaginaries framework is usefully read alongside the sociology of expectations, which examines how representations of future technologies shape present action \citep{Borup-2006-sociology, van-1993-promising}. Where Jasanoff and Kim are primarily interested in how imaginaries stabilize across institutions and publics, \citet{Borup-2006-sociology} are interested in the performative dynamics of expectation at closer range, how promising a technology creates the conditions (i.e., investment, talent, legitimacy) that make the promise more likely to be fulfilled. The two frameworks are complementary. The imaginary operates at the level of the collectively held vision and expectations operate at the level of the specific claims that build and sustain it. Both are relevant here because the essays analyzed do both things simultaneously. They project a civilizational future and they make specific technical claims whose function is partly to make the larger vision credible.

\citet{hilgartner-2015-science} developed the concept of the vanguard vision to describe the authoritative projection of a technological future by those closest to its development. The vanguard is not simply ahead of the curve, rather it claims the authority to define what the curve is (i.e., which futures are possible, which are desirable, and which are to be avoided). This authority is exercised through public performance like speeches, essays, demonstrations, and documents that constitute the vanguard as the legitimate interpreter of what the technology means. The two essays analyzed here are paradigmatic instances of vanguard vision production. Their authors are positioned at the frontier of AGI development and are widely granted the authority to speak on behalf of what is technically possible and what is humanly at stake. Crucially, this authority extends beyond the technical into the broadly social, because the imaginary the vanguard produces must address all the domains that transformative technology implicates.

\subsection{Corporate Imaginaries and the Retreat of the State}

\citet{Jasanoff-2015-Dreamscapes} extend the imaginaries framework to corporations, observing that private firms have become increasingly significant sites of imaginary production, particularly in technology sectors where the state has retreated from direct investment and governance. This extension raises a theoretical question the original framework does not fully resolve: what licenses treating corporate imaginaries through the same analytical lens as state ones, given that the sources of authority are so different? States derive their imaginary-producing authority from democratic legitimacy, territorial sovereignty, and ultimately coercive power. None of these apply to OpenAI or Anthropic. The extension therefore requires justification, and that justification is worth making explicit because it is also what makes the extension analytically productive rather than merely analogical.

The framework's portability rests not on the source of authority but on its function. What Jasanoff's account identifies as distinctive about sociotechnical imaginaries---that they stabilize a particular distribution of authority over technological futures by naturalizing it, and that they do so through institutionally backed public performance rather than explicit argument---applies to corporate imaginaries as fully as to state ones. The CEO essay addressed to humanity performs the same stabilizing work as the state science policy document addressed to citizens: it recruits a vision of the future into a collectively held understanding, fixes the terms on which that future is imaginable, and positions a specific institution as its legitimate steward. That the authority underwriting the performance derives from technical capacity and capital rather than from democratic mandate does not change the functional logic of the imaginary, on the contrary it changes the accountability structure within which that logic operates. And it is precisely this difference, the same stabilizing function exercised without the formal accountability that state actors bear, that is the paper's normative concern as much as its analytical one.

This configuration is not unprecedented. Scholars of American technology culture have traced how the mythology of the entrepreneurial innovator has historically served to naturalize the concentration of technical authority in private hands \citep{turner-2010-counterculture, mosco-2004-digital}. \citet{mosco-2004-digital} argues that digital technologies in particular have been systematically wrapped in what he calls the ``digital sublime'': a rhetoric of transcendence that presents technological development as the realization of deep historical and even spiritual necessity. This is continuous with a long tradition in which industry leaders have spoken on behalf of civilization-reshaping technologies \citep{nye-1996-american, hughes-1993-networks, noble-1979-america, jasanoff-1998-fifth, turner-2010-counterculture}, but the current configuration involves a specific change in the ecology of imaginary stabilization rather than merely its locus of production. Earlier digital-technology imaginaries were produced through a distributed network in which CEO voices were amplified and contested by journalists, critics, regulators, and a broader popular culture whose interpretive frames were not controlled by the firms themselves \citep{streeter-2010-net, turner-2010-counterculture, abbate-2000-inventing}. The essays examined here are produced at a moment when the gap between the technical community and the interpreting public is at its widest---where the capabilities of the systems being developed are opaque to almost everyone outside a small number of firms---and where that opacity concentrates the authority to define what the technology means more tightly than in earlier episodes. The significant update is not CEO authorship per se, but the degree to which the concentration of technical capacity in a small number of frontier labs has reduced the independent stabilizing infrastructure through which earlier technology imaginaries were contested and revised.

\citet{Ferretti-2021-Institutionalist} argues that AI governance has been excessively delegated to self-regulation not simply because firms are powerful but because they have successfully positioned themselves as the only actors with sufficient technical knowledge to govern responsibly. The essays analyzed in this paper are part of that positioning process. They do not merely reflect authority these firms already possess; they are part of the process through which that authority is produced and reproduced. Understanding the imaginary through Jasanoff's framework---extended to the corporate case on functional rather than analogical grounds---makes that process visible in a way that firm-level communication accounts do not.

A skeptical reader from a political economy or management background might reasonably interpret these essays as rational, equilibrium-signaling artifacts typical of capital-intensive, policy-sensitive markets: the rhetoric of inevitability maintains valuation, and the humanitarian framing manages regulatory risk. This reading is not wrong, but it is incomplete in a specific way. The rational-signaling account treats the imaginary as an instrument deployed by actors with pre-given interests, and treats those interests as independently legible. What the sociotechnical imaginaries framework adds is an account of how the interests themselves are constituted through the imaginary's production. The CEO who writes an essay about civilizational transformation is not merely signaling to investors; he is participating in the construction of the category of actor---the authorized steward of humanity's technological future---whose interests the essay then appears to express. The imaginary does not merely represent a pre-existing authority; it produces it. This distinction matters for governance: if the imaginary were merely a communication strategy layered over a set of given interests, disclosure requirements and PR reform might address it. If the imaginary is constitutive of the authority it appears to express, the governance challenge is structurally deeper.

\subsection{Imaginaries, Foreclosure, and Contestability}

A key insight of the imaginaries framework is that stabilized visions do not only describe possible futures but foreclose others. By fixing a particular account of what AGI will mean and how its development should proceed, an imaginary narrows what is imaginable as an alternative. Futures in which AGI development is governed differently, proceeds at a different pace, or is led by different institutions are not argued against. They simply do not appear. The foreclosure operates through the same naturalization that makes the proposed future feel inevitable. If the Intelligence Age or the compressed 21st century are simply the next stage in humanity's historical progression, then alternatives are not wrong but anachronistic.

This foreclosure has uneven distributional consequences. Scholarship on public perceptions of AI has consistently shown that non-expert publics, particularly in the global South, hold substantially different views of AI's risks and desirable futures than the technical communities producing it, views that tend to emphasize precaution, distributional equity, and local governance rather than the accelerationist optimism that characterizes both essays \citep{sartori-2023-minding, Jasanoff-2009-Containing}. Postcolonial STS scholars have further argued that the projection of a universalist technological future from a specific institutional and cultural position is itself a form of epistemic authority that requires critical examination \citep{visvanathan-1997-carnival, cath-2018-governing}. The populations who appear most prominently in both essays as beneficiaries of AGI-enabled progress, those suffering from disease, poverty, and authoritarianism in the developing world, do not participate in producing the imaginary that defines those futures. They appear as objects of the imaginary's beneficence rather than as its authors.

Understanding the imaginary as a mechanism of foreclosure rather than simply a vision of possibility redirects the critical question. Rather than asking whether the projections are accurate, it asks what institutional arrangements would be required to make the imaginary contestable from outside the firms that produce it. That question is taken up in the discussion.

\section{Data and Methods}
\label{sec:methods}

\subsection{Data}

This paper analyzes two documents: \citeauthor{altman-2024-intelligence}'s \textit{The Intelligence Age}, published on his personal website on September 23, 2024, and \citeauthor{amodei-2024-machines}'s \textit{Machines of Loving Grace}, published on his personal website in October 2024. Both were authored by the CEOs of the two leading AGI development firms, addressed to general audiences without prerequisite technical knowledge, and circulated widely across technical and popular media. Together they constitute the most sustained public articulations of each firm's vision of AGI's civilizational implications published to date.

The restriction to two documents is both a practical and an analytical choice. The CEO essay addressed to a general public is a distinct communicative genre with its own conventions and authority structure, and analyzing documents within the same genre controls for the variation that would be introduced by comparing, say, a shareholder letter with a congressional testimony or a technical blog post. More importantly, the comparative logic of the argument requires two documents that are comparable in type, proximate in time, and produced by institutions with meaningfully different cultures and stated risk philosophies. These conditions matter not because they rule out coordination, imitation, or shared market pressures as contributing explanations, and any of these factors may partly account for the convergence the analysis identifies, but because they make a purely mimetic or coordination-based account less plausible as a complete explanation. Two firms with different cultures, different risk philosophies, and leaders with notably different public personae, operating in direct competition with one another, face genuine incentives to differentiate their public imaginaries rather than replicate them. That they nevertheless produce the same underlying rhetorical architecture suggests that something beyond strategic choice or mutual imitation is doing explanatory work, that the imaginary reflects at minimum a feature of the institutional position these firms share, in addition to whatever firm-level or individual factors also contribute. These two documents satisfy the comparative requirements, and the structural claim they license is the stronger for being the more modest one.

The essays differ substantially in length and register. Altman's is approximately one thousand words and Amodei's is approximately ten thousand. Altman writes in the first person singular and in the declarative mood throughout. Amodei writes in a more hedged register, with frequent epistemic qualifiers and domain-specific elaboration. These differences are analytically productive rather than problematic, as they allow the analysis to distinguish between rhetorical style, which varies, and rhetorical structure, which does not.

\subsection{Methods}

The paper employs critical discourse analysis (CDA) and interpretive close reading. CDA attends to how texts construct social realities through specific linguistic and rhetorical choices, and is sensitive to what texts include and exclude, to the framing effects of particular terms, and to the relationship between textual claims and the institutional positions from which they are made \citep{Fairclough-1995-Critical}. Applied here, it interrogates not only what each essay claims but how those claims are structured, what rhetorical work specific moves perform, and what the essays collectively reveal about the conditions of imaginary production at these firms.

The analysis proceeds in two modes. Within-document analysis examines the internal rhetorical architecture of each essay, including its argumentative structure, its treatment of risk and uncertainty, and the narrative form through which the imaginary is constructed. Cross-document analysis examines what is shared across both essays despite differences in style and length, attending to the features stable enough to be attributed to structural position rather than individual choice. The interpretive logic is explicitly comparative, in that the unit of analysis is not either essay alone but the relationship between them.

A word on the limits of this method is warranted. CDA identifies the rhetorical operations through which texts construct meaning but does not establish how those texts are received. The analysis makes claims about what the essays do discursively, not about whether their audiences are persuaded, resist, or reinterpret them. Reception is a genuine and important question, one the conclusion identifies as a direction for future work, but it requires different methods and different data than those employed here. The present analysis is explicitly a production-side account. This methodological limit has a theoretical consequence the analysis must honor. The claim advanced here is more precisely that these essays constitute paradigmatic \textit{attempts} at imaginary projection, performances oriented toward stabilization rather than demonstrations that stabilization has been achieved. The analysis treats them as instantiations of an imaginary-in-formation rather than as evidence that the imaginary is fully stabilized in the Jasanoffian sense, which would require reception evidence this paper does not provide. Where the analysis uses the language of stabilization, it should be read as describing the rhetorical orientation of the texts, their design toward collective uptake, rather than as a claim about the uptake's success. Whether these visions have achieved the institutional embedding that full stabilization requires is an empirical question the conclusion correctly identifies as a direction for future work.

\section{The Imaginary in Performance: A Comparative Analysis}
\label{sec:analysis}

In September and October 2024, the CEOs of the two leading AGI development firms each published essays imagining the future that powerful AI would produce. Read together, the essays reveal that the utopian imaginary of AGI development is not an idiosyncratic feature of a single firm or a single leader's personality but a stable rhetorical structure deployed through meaningfully different stylistic approaches toward an identical ideological destination. This structural consistency is the paper's central empirical finding.

The essays differ substantially in length, register, and rhetorical identity. Altman's essay is roughly one thousand words, declarative in mood, and prophetic in register. Amodei's is roughly ten thousand words, hedged in its epistemic framing, and scientific in register. Altman performs the imaginary through compression while Amodei performs it through elaboration. Yet both arrive at the same place: AGI is historically inevitable, its arrival will be civilizationally transformative, and the firms developing it are indispensable to ensuring that transformation goes well. The four subsections that follow trace the specific rhetorical operations through which this shared destination is reached.

\subsection{The Self-Exemption Move}

Both essays open with a version of the same maneuver. They position their authors as uniquely exempt from the critique of AGI hype, and then proceed to enact it at length. This self-exemption move is analytically the most productive feature of either document, because it makes visible the mechanism the imaginary requires, the simultaneous disavowal and deployment of prophetic authority.

Amodei's version is the more elaborate and therefore the more revealing. Before the essay's main argument begins, he offers an extended account of why he and Anthropic have, until now, avoided discussing AI's upsides. The account is organized as a four-item list: he wants to ``maximize leverage'' by focusing on risks rather than benefits; he wants to ``avoid perception of propaganda,'' noting that ``AI companies talking about all the amazing benefits of AI can come off like propagandists, or as if they're attempting to distract from downsides''; he wants to ``avoid grandiosity,'' describing himself as ``often turned off by the way many AI risk public figures (not to mention AI company leaders) talk about the post-AGI world, as if it's their mission to single-handedly bring it about like a prophet leading their people to salvation''; and he wants to ``avoid sci-fi baggage,'' distancing himself from futures that feature ``uploaded minds, space exploration, or general cyberpunk vibes'' \citep{amodei-2024-machines}. The list is unusually precise in its self-diagnosis. Amodei names the pathology---prophetic grandiosity, propagandistic intent, subcultural fantasy---and distances himself from each in turn.

The gap between this framing section and the essay that follows is the document's most important feature. The essay goes on to project the cure of most infectious disease, the elimination of most cancer, the prevention of Alzheimer's, the doubling of the human lifespan, the cure of most mental illness, the lifting of the developing world to current US income levels, and what Amodei calls ``eternal 1991''---a permanent global triumph of liberal democracy \citep{amodei-2024-machines}. The essay that was introduced as a corrective to prophetic grandiosity is itself an act of prophetic grandiosity on a civilizational scale. Amodei has diagnosed the structure of the utopian imaginary, explicitly rejected it on principled grounds, and then enacted it across forty-five pages.

This is not evidence of bad faith. It is evidence of how structurally powerful the imaginary is. The self-exemption move does not free Amodei from the imaginary; it deepens his investment in it by adding a layer of apparent self-awareness. Consider the specific language of the grandiosity bullet: he is ``turned off'' by leaders who approach development ``as if it's their mission to single-handedly bring it about like a prophet leading their people to salvation,'' and describes such thinking as ``dangerous'' and ``quasi-religious'' \citep{amodei-2024-machines}. The critique is leveled at a third-person ``they''---other AI leaders, unnamed---while Amodei positions himself as the observer who has seen through the pathology. But the essay he then writes is addressed to humanity, projects civilizational transformation, and ends with the sentence: ``We have the opportunity to play some small role in making it real'' \citep{amodei-2024-machines}. The modesty of ``some small role'' is syntactically present but rhetorically cancelled by everything that precedes it. Far from undermining the imaginary's authority, the critique of prophetic grandiosity becomes one more performance of reasonableness that makes the imaginary more credible to its audience.

The ``sci-fi baggage'' bullet is equally revealing under scrutiny. Amodei objects not to the content of utopian futures but to their ``vibe''---the subcultural connotations that cause people to take such visions less seriously. The problem with sci-fi framings, he writes, is that they ``connotatively smuggle in a bunch of cultural baggage and unstated assumptions about what kind of future is desirable'' \citep{amodei-2024-machines}. This is a critique of rhetorical packaging, not of the underlying epistemic move. Amodei's solution is to deliver the same civilizational projections in a different register: not uploaded minds but doubled lifespans; not space exploration but compressed biological progress. The content is no less radical---he acknowledges this himself, noting that ``my predictions are going to be radical as judged by most standards (other than sci-fi singularity visions)'' \citep{amodei-2024-machines}---but the packaging is clinical rather than speculative. The self-exemption move and the genre critique together perform the same function: they clear a space in which Amodei can produce the imaginary while appearing to stand outside it.

Altman's self-exemption is far briefer but structurally identical. His single concession to risk reads: ``The dawn of the Intelligence Age is a momentous development with very complex and extremely high-stakes challenges. It will not be an entirely positive story, but the upside is so tremendous that we owe it to ourselves, and the future, to figure out how to navigate the risks in front of us'' \citep{altman-2024-intelligence}. The concession occupies one sentence. The imaginary reasserts itself in the next. The rhetorical function is the same as Amodei's elaborated version: the acknowledgment of complexity establishes the author's credibility as a serious thinker, while the pivot to ``the upside is so tremendous'' absorbs the acknowledgment into the optimistic frame before it can do any critical work. Altman's brevity here is itself a rhetorical choice: the prophet does not linger on doubt.

What the self-exemption move reveals, across both essays, is that the imaginary is not an individual rhetorical choice but a structural compulsion. \citeauthor{Jasanoff-2015-Dreamscapes}'s account of imaginary stabilization explains why: once an imaginary achieves institutional backing and public circulation, individual actors are constrained by it even when they are its producers \citep{Jasanoff-2015-Dreamscapes}. Amodei cannot simply opt out of the genre that gives his essay its authority. The CEO essay about AGI's civilizational consequences has its own rhetorical logic, and that logic requires the utopian projection regardless of the author's stated misgivings. The self-exemption move does not escape this logic; it confirms it.

A rival interpretation would read Amodei's framing section as ordinary genre-conventional hedging---the kind of credibility-building prebuttal that any careful writer deploys before making strong claims. The difference between self-exemption and standard hedging, however, is structural rather than tonal. Standard hedging qualifies the specific claims that follow; the self-exemption move here quarantines an entire category of criticism---prophetic grandiosity, propagandistic intent---and then proceeds to enact it in full. The test is not whether the author acknowledges uncertainty but whether the acknowledgment does any structural work: whether it narrows the scope of subsequent claims, introduces genuine alternative possibilities, or survives contact with the essay that follows. In Amodei's case, the framing section performs none of these functions. The forty-five pages that follow are not narrower in scope, more genuinely uncertain, or more alternative-inclusive than they would have been without the preface. The self-exemption is diagnostically distinct from hedging precisely because its effect is to expand rather than constrain the authority of what follows.

\subsection{Teleological Naturalization and the Inevitability of AGI}

Both essays embed AGI development in teleological narratives of human progress that naturalize its arrival as historically inevitable. This naturalization is the imaginary's foundational operation: it transforms a contingent set of technical and commercial decisions---made by specific firms, funded by specific investors, governed by specific incentive structures---into the expression of an underlying historical logic that no one chose and no one can reasonably oppose. The mechanism operates differently in each essay, but the ideological effect is the same.

Altman's teleology is sequential and civilizational. He opens with intergenerational continuity---``in the next couple of decades, we will be able to do things that would have seemed like magic to our grandparents''---before ascending to a five-word historical arc: ``Technology brought us from the Stone Age to the Agricultural Age and then to the Industrial Age. From here, the path to the Intelligence Age is paved with compute, energy, and human will'' \citep{altman-2024-intelligence}. The compression of human history into a sequence of named Ages performs naturalization at the level of grammar: AGI is not a contingent outcome of recent investment decisions but the next member of a series whose logic has been unfolding for millennia. The phrase ``the path to the Intelligence Age'' is particularly important. A path implies that the destination exists prior to and independent of any traveler's choice to walk it; Altman is not building a future but disclosing one that was already there.

The ``melt sand'' passage performs the same naturalization at a finer grain and is worth examining at the word level. Altman writes: ``after thousands of years of compounding scientific discovery and technological progress, we have figured out how to melt sand, add some impurities, arrange it with astonishing precision at extraordinarily tiny scale into computer chips, run energy through it, and end up with systems capable of creating increasingly capable artificial intelligence'' \citep{altman-2024-intelligence}. The passage is structured as a single cumulative sentence whose subject is the indefinite ``we'' of all of humanity across all of history. The verb ``have figured out'' locates the discovery at the end of a long process of compounding rather than at a specific moment of investment or institutional decision. The choice of ``melt sand'' as the entry point is rhetorically precise: it begins at the material substrate, as far upstream as possible, so that the passage from raw earth to artificial intelligence appears as a natural unfolding rather than a series of choices. AGI's development is presented as something humanity arrived at the way water arrives at the sea---not by deciding to go there but by following the logic of its own accumulation. The phrase ``this may turn out to be the most consequential fact about all of history so far'' \citep{altman-2024-intelligence} then caps the teleology: not the most consequential decision, or the most consequential investment, but the most consequential \textit{fact}---a thing that simply is, whose existence is independent of any actor's agency.

The compression of the entire technical research program into three words---``deep learning worked'' \citep{altman-2024-intelligence}---performs the same naturalization in its most extreme form. An entire history of contested technical decisions, competing research paradigms, enormous capital allocation, and institutional competition is compressed into a past-tense declaration that reads as brute fact. The passive construction---it worked, rather than we made it work, or we bet that it would work---strips the development of its contingency. It did not work because specific firms made specific choices; it worked the way gravity works.

Amodei's teleology is philosophically more ambitious and, at the essay's conclusion, more explicit about its ideological stakes. He does not merely argue that AGI will accelerate progress; he argues that it will vindicate the direction in which humanity was already headed. The essay's final pages invoke Iain M. Banks' \textit{The Player of Games}, in which a protagonist from a society based on liberal democratic values defeats an emperor whose game was designed to reward ruthless competition, to argue that the Culture's values ``represent a winning strategy even in a game designed by a society based on ruthless competition and survival of the fittest'' \citep{amodei-2024-machines}. The political argument is made through literary analogy: liberal democracy wins not because it is powerful but because it is, in some deep sense, correct---because it represents the logical endpoint of basic human moral intuitions.

Amodei then states the teleological claim directly: ``These simple intuitions, if taken to their logical conclusion, lead eventually to rule of law, democracy, and Enlightenment values. If not inevitably, then at least as a statistical tendency, this is where humanity was already headed. AI simply offers an opportunity to get us there more quickly---to make the logic starker and the destination clearer'' \citep{amodei-2024-machines}. Three moves are packed into these sentences. First, the current political order---liberal democracy, Enlightenment values, rule of law---is presented not as a contingent historical outcome but as the logical endpoint of universal moral intuitions. Second, this endpoint is described as where humanity ``was already headed''---AGI does not create the direction, it accelerates a trajectory that already existed. Third, AGI is positioned as the agent that makes ``the logic starker and the destination clearer,'' which means that opposition to AGI development is not merely technical skepticism but resistance to moral clarity. Where Altman's teleology frames opposition as anachronistic, Amodei's frames it as something closer to moral error.

The phrase ``eternal 1991'' condenses this political teleology into its most charged form \citep{amodei-2024-machines}. The reference is to Fukuyama's end-of-history thesis, which Amodei both invokes and refines: where Fukuyama's claim was that liberal democracy had prevailed as a matter of historical fact, Amodei's is that AGI will ensure it prevails as a matter of permanent structural advantage. ``Eternal'' is doing significant ideological work here---it forecloses the historical contestation that the rest of the governance section acknowledges as real and ongoing. The essay that has spent several pages noting the fragility of democracy, the threat of authoritarian AI, and the difficulty of the ``fight'' for liberal values ends by projecting those values as eternal. The concession to fragility is absorbed by the imaginary's final move: what is fragile becomes permanent once the right technology is in place.

\subsection{Qualified Acknowledgment and Implicit Indispensability}

Both essays treat risk and commercial interest through a pair of related rhetorical operations that are best analyzed together, because they perform complementary functions in naturalizing the firms' authority. Qualified acknowledgment absorbs genuine concessions to risk into the optimistic frame; implicit indispensability establishes the firm as central to the imagined future without naming it as a commercial actor. Together they constitute the imaginary's defensive architecture.

The pattern of qualified acknowledgment is structurally consistent across both essays: here is a real constraint, here is why it will be overcome, therefore the optimistic projection holds. Every acknowledged risk becomes a speed bump rather than a stop sign. In Altman's essay this pattern is performed at a high level of abstraction. Labor displacement receives a single paragraph in which the concern is named and immediately reframed: ``most jobs will change more slowly than most people think, and I have no fear that we'll run out of things to do'' \citep{altman-2024-intelligence}. The lamplighter analogy that follows---no one looks back wishing they were a lamplighter; if a lamplighter could see today's world he would find the prosperity unimaginable---performs a temporal displacement that makes the concern feel provincial. The reader is positioned as someone who, like the lamplighter, cannot yet see the world that is coming; to worry about job displacement is to be trapped in a limited historical vantage point.

In Amodei's essay, risk is engaged more substantively through the factors-of-production framework, which genuinely attempts to bound the imaginary with technical reasoning. The five limiting factors---the speed of the physical world, the need for data, intrinsic complexity, constraints from humans, and physical laws---represent a real analytical effort to identify where intelligence reaches diminishing returns. But the framework's rhetorical function is to make the optimistic projections appear derived from careful reasoning rather than presupposed by it. The numbers that punctuate the biology and neuroscience sections---``at least 10x,'' ``maybe 1000x,'' ``50--100 years of progress in 5--10 years''---are presented with the syntax of calculation while resting on assumptions Amodei himself describes as speculative. He writes: ``it's my guess that powerful AI could at least 10x the rate of these discoveries, giving us the next 50--100 years of biological progress in 5--10 years. Why not 100x? Perhaps it is possible\ldots I'm actually open to the (perhaps absurd-sounding) idea that we could get 1000 years of progress in 5--10 years'' \citep{amodei-2024-machines}. The hedge ``perhaps absurd-sounding'' is present but it is immediately bracketed by the willingness to entertain it. The numerical range (10x to 1000x, a two-order-of-magnitude spread) is not a product of a model; it is a rhetorical anchor that makes the imaginary feel calculated. \citet{Borup-2006-sociology} observe that quantified expectations perform a specific kind of epistemic authority in technoscientific discourse---they invite engagement on technical terms while foreclosing the prior question of whether the framework generating the numbers is sound. Amodei's figures work exactly this way.

The operation of implicit indispensability is a shared feature of both essays, and it works through absence rather than assertion. Neither essay mentions competitors, market share, revenue projections, investor obligations, or the commercial dynamics of AGI development. But the absence of explicit self-promotion does not mean the firms' centrality goes unestablished; it means that centrality is established through other means. Five mechanisms are identifiable across both texts: the collective ``we'' that slides between humanity and the firm without marking the transition; the problem-solution structure that makes the firms' technical program appear necessary to address civilizational challenges; the absence of named alternatives, which makes the firms' role seem structural rather than chosen; credentialed proximity, in which the authority to make projections and the indispensability those projections establish are mutually constituting; and the humanitarian pivot, which frames the cost of not proceeding as the continued suffering of the world's most vulnerable populations, shifting the burden of justification onto critics rather than the firms. The close reading that follows traces the first three of these mechanisms, which are most visible in the texts; the latter two operate more diffusely across both essays as a whole.

Anthropic appears twice in Amodei's essay: once in the opening sentence, identifying the author's institutional affiliation, and once in a passing reference to a ``computational democracy project, including collaborations with Anthropic'' \citep{amodei-2024-machines}. The restraint is continuous with the self-exemption strategy: Amodei has explained that he wants to avoid the appearance of propaganda, so Anthropic remains offstage. But the essay's existence, its author's identity, and its institutional circulation ensure that Anthropic is the primary beneficiary of the humanitarian imaginary it projects. The firm that builds the AI capable of doubling human lifespan, curing most mental illness, and ensuring the permanent triumph of liberal democracy is indispensable by construction---even if the essay never names it as such.

This is what distinguishes the corporate imaginary from the state imaginaries that \citeauthor{Jasanoff-2015-Dreamscapes} originally analyzed. State actors must openly claim public authority; their legitimacy depends on making that claim visible and subjecting it to democratic contestation. Private firms can establish authority through the imaginary itself, by defining the future in terms that make their own role appear structurally necessary rather than commercially motivated. The humanitarian framing is not a mask worn over a commercial interest; it is the mechanism through which the commercial interest is naturalized as a public one. The firms do not need to say they are indispensable; they need only describe a future that cannot be reached without them.

\subsection{The Structural Finding}

Read together, the most significant conclusion is not any particular claim either essay makes but the structural similarity of how both essays work. Two documents written by the leaders of competing firms, published within weeks of each other, deploy the same underlying architecture: the self-exemption move that disavows prophetic authority while exercising it; the teleological narrative that makes AGI's arrival appear as the next stage in humanity's historical or moral progression; the qualified acknowledgment that gives concessions to risk no structural weight; and the implicit indispensability that positions the firm as central to the imagined future without naming it as a commercial actor.

The rhetorical execution is meaningfully different. Altman is a prophet; Amodei is a scientist. Altman compresses; Amodei elaborates. But the ideological structure is identical, and this identity is the section's key finding. If the imaginary were merely a reflection of individual personality or firm-level communication strategy, we would expect greater variation between two leaders with such different public personae and such different rhetorical styles. The competitive incentives point requires qualification: in high-uncertainty, nascent fields, convergence on a common future narrative can itself be strategically rational---through mimetic isomorphism, shared category legitimation, or coordinated expectation-building that raises all boats. The argument here does not assume that competition must produce divergence. Rather, it is that the convergence is at the level of rhetorical \textit{architecture} rather than surface style---that it survives precisely the variation one would expect from strategic differentiation across two firms with genuinely different risk philosophies, different stated positions on the pace of development, and leaders whose public personae are in meaningful tension. It is this depth of convergence, not the mere fact of it, that makes a purely strategic or mimetic account less plausible as a complete explanation. The fact that the imaginary is structurally consistent across both suggests it is not \textit{only} a choice but \textit{also} a compulsion---a feature of the institutional position these firms occupy that operates alongside, and cannot be reduced to, the preferences of the individuals who lead them or the strategic communication choices those individuals make.

Firms developing transformative technologies at the frontier of what is technically possible, in the absence of established regulatory frameworks, with enormous commercial stakes and significant public scrutiny of their motives, are drawn toward this imaginary because it performs the essential work their position requires: it naturalizes their authority, establishes their indispensability, and forecloses the space in which alternative futures might be imagined. This gravitational pull does not mean that individual choices, firm cultures, or competitive dynamics are irrelevant---they shape the execution, as the meaningful differences between the two essays demonstrate. But it does mean that those factors alone are insufficient to explain the convergence. The imaginary, at minimum, reflects something structural about the position these firms occupy, and that structural dimension is what makes it likely to persist across firms, leaders, and communications strategies that may otherwise differ substantially. \citet{Jasanoff-2015-Dreamscapes} characterize sociotechnical imaginaries as encoding not only visions of what is attainable through science and technology but also of ``how life ought, or ought not, to be lived''---a formulation that captures the normative load the essays carry beyond their technical claims. The essays examined here are not merely descriptions of possible futures. They are performances of power---institutionally backed, publicly staged, and oriented toward securing the authority of the firms that produce them to define what the future will look like and who will lead humanity there.

The structural claim advanced here is deliberately scoped. It does not assert that any firm developing powerful AI must produce this imaginary, but rather that firms occupying a specific configuration of conditions---safety-maximalist orientation, policy-facing public posture, frontier technical position, and civilizational stakes claims---face compulsions toward this rhetorical architecture that cannot be fully explained by coordination or imitation alone. The absence of a directly comparable negative case is itself informative: no other firm currently occupies this precise configuration. This is a limitation of the analysis, but it also narrows the scope of the claim in a way that makes it more defensible rather than less. The structural argument is not that all AI firms converge on this imaginary, but that the specific position OpenAI and Anthropic share generates compulsions that shape even leaders with strong incentives and stated commitments to differentiate.

\section{Authority, Foreclosure, and the Question of Contestability}
\label{sec:discussion}

The analysis demonstrates that OpenAI and Anthropic construct sociotechnical imaginaries through a structurally consistent strategy. Their CEOs' public essays naturalize AGI development as historically and morally inevitable, establish the firms' indispensability without naming them as commercial actors, and absorb genuine concessions to risk into an overarching optimistic frame---all while performing distinct rhetorical identities that make the shared imaginary persuasive to different audiences. This section draws out three implications that extend beyond the texts themselves.

\subsection{Private Firms as Primary Imaginary Architects}

The most significant implication of the analysis concerns the institutional location of imaginary production. \citet{Jasanoff-2015-Dreamscapes} developed the sociotechnical imaginaries framework primarily through comparative analysis of nation-states. When the US government promoted nuclear power through the ``Atoms for Peace'' program, or when South Korea constructed its developmental state imaginary around nuclear energy \citep{Jasanoff-2009-Containing}, the institution producing the imaginary was the state, with the formal accountability and democratic legitimacy that implies. The institution that defines the technological future is, on this model, also the institution answerable to the public whose future it is defining.

OpenAI and Anthropic occupy no equivalent position of formal accountability. They are private firms with fiduciary obligations to investors and, in OpenAI's case, a governance structure that has already demonstrated its fragility under commercial pressure. Yet they have become the primary architects of the sociotechnical imaginary through which AGI's futures are publicly understood---not because they have claimed this position illegitimately, but because the technical capacity to develop powerful AI is concentrated in them, and with technical capacity comes the authority to define what the technology means. The essays analyzed here are not merely products of that authority; they are part of the process through which it is produced and reproduced.

This concentration of imaginary-making power in private hands is not unprecedented in American technology culture. \citet{turner-2010-counterculture} traces how the mythology of the entrepreneurial innovator---the visionary whose individual genius reshapes civilization---has historically naturalized the concentration of technical authority in private institutions, framing what are essentially commercial decisions as acts of humanistic beneficence. \citet{mosco-2004-digital} identifies a recurring structure he calls the digital sublime: a rhetoric of transcendence that presents technological development as the realization of deep historical necessity, obscuring the specific institutional interests that benefit from its deployment. Both essays analyzed here are continuous with this tradition. What is new is the scale of the stakes and the explicitness of the civilizational claim. Earlier instances of the digital sublime were produced diffusely, across journalism, marketing, and popular culture. The essays examined here are produced by the CEOs of the firms doing the technical work, which consolidates in a single institutional location the authority to build the technology and the authority to define what it means.

This consolidation has a specific historical condition: the retreat of the state from direct investment in and governance of frontier technology development. \citet{mazzucato-2013-entrepreneurial} argues that the technologies undergirding the current AI moment---the internet, GPS, touchscreens, the foundational research enabling deep learning itself---were substantially products of public investment, yet their commercial exploitation has been captured almost entirely by private firms. The imaginaries these firms now produce present AGI development as a private achievement whose civilizational significance licenses private authority over its governance. The state that created the conditions for the technology does not appear in either essay as a legitimate shaper of its future; it appears, in Amodei's ``entente strategy'' \citep{amodei-2024-machines}, primarily as a military and geopolitical instrument that AI firms should seek to align with their interests. \citet{jasanoff-2018-global} argue that what is needed instead is a form of ``technology statecraft''---a recovery of public deliberative capacity adequate to the governance of transformative technologies. The essays analyzed here are evidence of precisely the gap that such statecraft would need to fill.

\subsection{Foreclosure and the Costs of the Imaginary}

Both essays address ``humanity'' in the aggregate, but the imaginary they produce is not universal in the sense of being produced by or for everyone. It is produced from a specific institutional and cultural position---the American technology industry at the frontier of AGI development---and it encodes the assumptions and priorities of that position, including assumptions about which risks are most salient, which futures are worth building toward, and who counts as a relevant stakeholder.

The populations who appear most prominently in both essays are those in the developing world suffering from disease, poverty, and authoritarian governance. Amodei projects that AI-enabled growth rates could bring sub-Saharan Africa to current Chinese per-capita GDP levels within a decade \citep{amodei-2024-machines}; Altman promises that AI will be put ``into the hands of as many people as possible'' \citep{altman-2024-intelligence}. These populations are the imaginary's most prominent beneficiaries. They are not, however, its producers. They do not participate in defining what AGI should be, what risks are acceptable, or what futures are worth building toward. They appear as objects of the imaginary's beneficence---as the humanitarian justification for a development agenda they had no role in shaping.

This exclusion is structural rather than incidental, and its consequences extend beyond the unfairness of non-participation. \citet{winner-1980-artifacts} argues that technological arrangements are not politically neutral: they embed specific distributions of authority, access, and risk that persist long after the moment of design. The imaginary produced by these essays shapes the terms on which AGI development proceeds---what is built, at what pace, governed by whom, for whose benefit---in ways that will be very difficult to contest once the imaginary is stabilized. Postcolonial STS scholars have argued that the projection of a universalist technological future from a specific institutional and cultural position is itself an exercise of epistemic authority that requires critical examination \citep{ndlovu-gatsheni-2018-epistemic, birhane-2021-algorithmic}. The populations whose futures are most consequentially shaped by AGI development do not participate in producing the imaginary that defines those futures; they are, in \citeauthor{ndlovu-gatsheni-2018-epistemic}'s terms, epistemic objects of the imaginary rather than its subjects.

The foreclosure operates not only across geographies but across possible futures. By stabilizing a particular vision of what AGI will mean and how its development should proceed, the essays narrow what is imaginable as an alternative. Futures in which AGI development is governed differently, proceeds at a different pace, serves different populations, or is led by different institutions are not argued against; they simply do not appear. The naturalization of AGI's inevitability performs this foreclosure preemptively: if the Intelligence Age or the compressed 21st century are simply the next stage in humanity's historical progression, then alternatives are not wrong but anachronistic. \citet{couldry-2019-costs} argue that the most considerate feature of data capitalism is not the extraction of value from user data but the foreclosure of alternative ways of organizing social and economic life that data-driven systems make progressively harder to imagine. The imaginaries analyzed here operate by a similar logic: they do not simply describe a future, they narrow the conceptual space in which other futures might be proposed and defended.

\subsection{Toward Contestability}

Understanding the imaginary as structural rather than strategic has direct implications for how we think about governing AGI development. If the imaginary were merely a communication strategy---a set of claims firms could be persuaded or required to make differently---then governance interventions could focus on disclosure requirements, risk representation standards, or stakeholder consultation mandates. These are not without value, but they address the surface of the problem rather than its structure. If the imaginary is produced by the logic of the institutional position rather than by deliberate rhetorical choice, then such reforms will encounter resistance not merely from individual firms but from the compulsions that produce the imaginary in the first place.

The normative recommendations that follow are keyed to the specific discursive operations identified in the analysis, because a governance response that does not address the mechanism is unlikely to affect it. If the problem diagnosed by teleological naturalization is the presentation of contingent decisions as historical necessity, the relevant governance response is not generic transparency but \textit{denaturalization}: institutional forums in which the contingency of the development trajectory is made explicit and the available branching points are named as choices rather than facts. If the problem diagnosed by implicit indispensability is the naturalization of private commercial authority as public beneficence, the relevant mechanism is not disclosure of commercial interest per se but \textit{forced appearance as a commercial actor}---regulatory or deliberative contexts in which the firms must present their interests as interests, subject to contestation on equal terms with other stakeholders' interests, rather than as the expression of humanity's historical progress. Generic calls for humility are inadequate precisely because the imaginary has, as the analysis shows, already inoculated itself against such calls through the self-exemption move: the firms have performed humility as a precondition of exercising authority. What is needed instead are institutional arrangements that do not merely invite humility but structurally require the appearance of the commercial actor---that make the self-exemption move unavailable rather than merely visible.

The deeper problem is what \citet{jasanoff-2003-technologies} calls the deficit of ``technologies of humility''---institutionalized practices that make the assumptions embedded in technological development visible and contestable before they are stabilized. The essays analyzed here are striking precisely because they make their assumptions invisible through the rhetorical operations identified in Section~\ref{sec:analysis}: the teleological narrative that presents contingent decisions as historical necessity, the self-exemption that performs reasonableness while foreclosing scrutiny, the absence of commercial self-reference that naturalizes private authority as public beneficence. Technologies of humility would work in the opposite direction: they would create forums in which the assumptions embedded in the imaginary are named, the alternatives it forecloses are made visible, and the populations it addresses as beneficiaries are able to participate as authors.

What such forums would look like in practice is a genuinely difficult question, and this paper does not attempt to answer it fully. The specific features of AGI development that complicate standard deliberative mechanisms are not merely speed and technical complexity, though both are real obstacles. The deeper difficulty is epistemic: the capabilities of the systems being developed are genuinely uncertain even to the people developing them. When the developers themselves cannot specify what they are building with precision, participatory technology assessment models designed for more tractable domains---genomics, nuclear waste management, infrastructure planning---do not straightforwardly apply. Deliberation about tradeoffs presupposes that the tradeoffs are knowable in advance; at the AGI frontier, that presupposition fails. This is not an argument against deliberative institutions but an argument for designing them with that epistemic condition in mind, rather than importing models from domains where it does not obtain.

What this paper can contribute to that design project is more limited but not trivial: it makes the imaginary's construction visible. Showing that what presents itself as prophecy or scientific assessment is a specific rhetorical architecture with identifiable operations and identifiable effects is a necessary condition for contestability, even if it is not a sufficient one. An imaginary that appears natural---that presents AGI's arrival as the next stage in humanity's historical progression and the firms developing it as its necessary stewards---is very difficult to contest on its own terms, because its terms are the ones in which the debate is conducted. An imaginary whose construction has been made visible is at least contestable in principle: its assumptions can be named, its foreclosures identified, and its authority questioned from outside the institutions that produce it. That is what the analysis offered here attempts. The further institutional question---what forums, with what composition, operating by what procedures, would be adequate to the specific epistemic conditions of AGI development---is the most important question the findings raise, and it is one that requires collaborative work across STS, deliberative democratic theory, and AI governance scholarship that this paper cannot complete alone.

The point is not that Altman and Amodei are wrong about what AGI will mean or wrong to want to communicate their vision publicly. The point is that their visions, however thoughtfully constructed, are not sufficient substitutes for broader deliberation about the futures that transformative technology makes possible and forecloses. The essays analyzed here are, in the end, performances of authority---claims about who is qualified to define the future and on what terms. Making that authority contestable does not require silencing the firms that produce it. It requires, first, making the construction of that authority visible; and second, creating institutional conditions adequate to the specific epistemic challenges of deliberating about a technology whose capabilities remain uncertain even to its developers. This paper addresses the first requirement. The second remains open.

\section{Conclusion}
\label{sec:conclusion}

This paper has argued that OpenAI and Anthropic construct sociotechnical imaginaries of AGI through a structurally consistent rhetorical strategy, despite meaningful differences in execution. The self-exemption move disavows prophetic authority while exercising it; teleological naturalization embeds contingent commercial decisions in narratives of historical and moral inevitability; qualified acknowledgment gives concessions to risk no structural weight; and implicit indispensability positions each firm as central to the imagined future without naming it as a commercial actor with competitive interests. That two institutions with different cultures, different risk philosophies, and leaders with notably different public personae independently produce the same underlying architecture---despite strong competitive incentives to differentiate their public visions---is the paper's central empirical finding. The imaginary is not only a firm-level communication choice. It is, at minimum, a feature of the position these firms occupy.

The paper makes three contributions to the sociotechnical imaginaries literature. First, it extends the framework from nation-states to private firms at the frontier of transformative technology development, justifying that extension on functional rather than analogical grounds: corporate and state imaginaries perform the same stabilizing work through institutionally backed public performance, and the difference in the source of their authority is precisely what generates the paper's normative concern rather than undermining the framework's applicability. Second, the paper identifies the specific rhetorical operations through which corporate imaginaries are constructed: the four operations analyzed in Section~\ref{sec:analysis} are analytically distinguishable, together constitute a mechanism, and can be identified and examined in other firms and other technologies. Third, the comparative method demonstrates that the mechanism is at minimum structural rather than purely idiosyncratic, providing a foundation for the systematic comparative research the field needs.

Several directions for future work follow from this analysis. A broader comparative study examining the imaginaries produced by other leading AI firms---Google DeepMind, Meta AI, xAI---would test whether the mechanism identified here is specific to the two firms analyzed or characteristic of the sector as a whole. Longitudinal analysis tracking how these firms' imaginaries evolve in response to regulatory pressure, public controversy, and technological developments would illuminate the temporal dynamics of imaginary stabilization and contestation. Reception studies examining how different publics interpret and respond to these imaginaries---particularly the populations in the global South whose futures the essays invoke but whose voices they do not include---would complement the production-focused analysis offered here and test whether the foreclosure the imaginary performs is as complete in practice as it appears in the texts. Most pressingly, collaborative work across STS, deliberative democratic theory, and AI governance scholarship is needed to specify what institutional forums, with what composition and what procedures, would be adequate to the specific epistemic conditions of AGI development---where the capabilities of the systems being built remain uncertain even to their developers.

The broader stakes of this analysis are not primarily textual. The essays examined here are performances of authority at a moment when the question of who is authorized to define AGI's future remains genuinely open, and they have a structural advantage in the contest over that question: they are produced by the institutions with the greatest technical capacity, the most resources, and the most direct access to the public forums in which the imaginary is stabilized. Contesting that authority requires, first, making its construction visible---showing that what presents itself as prophecy or scientific assessment is a specific rhetorical architecture with identifiable operations and identifiable effects. That is what this paper has attempted. The second requirement is institutional, and it remains open.

Amodei ends his essay with a sentence worth sitting with: ``We have the opportunity to play some small role in making it real'' \citep{amodei-2024-machines}. The modesty is syntactically present---``some small role''---and rhetorically cancelled by everything that precedes it. A ten-thousand-word essay projecting the permanent triumph of liberal democracy, the doubling of the human lifespan, and the cure of most human disease does not end in a minor key. The sentence performs the same self-exemption the essay opened with: it claims humility at the moment of maximum ambition. But the sentence also reveals something the imaginary cannot entirely conceal: that the future it projects is not yet real, that making it real will require choices, and that choices can be made differently. The imaginary presents those choices as already settled by history. The task for scholars, policymakers, and publics is to insist that they are not.

\bibliographystyle{plainnat} 
\bibliography{refs}

@misc{altman-2024-intelligence,
  author       = {Altman, Sam},
  title        = {The Intelligence Age},
  year         = {2024},
  month        = sep,
  day          = {23},
}

@misc{amodei-2024-machines,
  author       = {Amodei, Dario},
  title        = {Machines of Loving Grace: How AI Could Transform the World for the Better},
  year         = {2024},
  month        = oct,
  day          = {16},
}

@book{Jasanoff-2015-Dreamscapes,
	author = {Jasanoff, Sheila and Kim, Sang-Hyun},
	doi = {10.7208/chicago/9780226276663.001.0001},
	isbn = {9780226276526},
	year = {2015},
	title = {Dreamscapes of {Modernity}},
    publisher = {The University of Chicago Press},
}

@article{Jasanoff-2009-Containing,
	author = {Jasanoff, Sheila and Kim, Sang-Hyun},
	journal = {Minerva},
	doi = {10.1007/s11024-009-9124-4},
	issn = {0026-4695},
	number = {2},
	year = {2009},
	month = {6},
	pages = {119--146},
	publisher = {{Springer Science and Business Media LLC}},
	title = {Containing the {Atom}: Sociotechnical {Imaginaries} and {Nuclear} {Power} in the {United} {States} and {South} {Korea}},
	volume = {47},
}

@article{sartori-2023-minding,
  title={Minding the gap (s): public perceptions of AI and socio-technical imaginaries},
  author={Sartori, Laura and Bocca, Giulia},
  journal={AI \& society},
  volume={38},
  number={2},
  pages={443--458},
  year={2023},
  publisher={Springer}
}

@book{hilgartner-2015-science,
  title={Science and democracy},
  author={Hilgartner, Stephen and Hagendijk, Rob and Miller, Clark},
  year={2015},
  publisher={Routledge}
}

@book{fairclough-1995-critical,
  title={Critical discourse analysis: The critical study of language},
  author={Fairclough, Norman},
  year={1995},
  publisher={Routledge}
}

@article{Ferretti-2021-Institutionalist,
	author = {Ferretti, Thomas},
	journal = {Moral Philosophy and Politics},
	doi = {10.1515/mopp-2020-0056},
	issn = {2194-5616},
	number = {2},
	year = {2021},
	month = {aug 5},
	pages = {239--265},
	publisher = {Walter de Gruyter GmbH},
	title = {An {Institutionalist} {Approach} to {AI} {Ethics}: Justifying the {Priority} of {Government} {Regulation} over {Self}-{Regulation}},
	volume = {9},
}

@article{van-1993-promising,
  title={Promising Technology. The Dynamics of Expectations in Technological Developments},
  author={van Lente, H},
  year={1993}
}

@article{Borup-2006-sociology,
	author = {Borup, Mads and Brown, Nik and Konrad, Kornelia and Van Lente, Harro},
	journal = {Technology Analysis \&amp; Strategic Management},
	doi = {10.1080/09537320600777002},
	issn = {0953-7325},
	number = {3-4},
	year = {2006},
	month = {7},
	pages = {285--298},
	publisher = {Informa UK Limited},
	title = {The sociology of expectations in science and technology},
	volume = {18},
}

@book{turner-2010-counterculture,
  title={From counterculture to cyberculture: Stewart Brand, the Whole Earth Network, and the rise of digital utopianism},
  author={Turner, Fred},
  year={2010},
  publisher={University of Chicago Press}
}

@article{mosco-2004-digital,
  title={The digital sublime: myth, power, and cyberspace MIT Press},
  author={Mosco, V},
  journal={Cambridge MA},
  year={2004}
}

@book{visvanathan-1997-carnival,
  author    = {Visvanathan, Shiv},
  title     = {A Carnival for Science: Essays on Science, Technology and Development},
  year      = {1997},
  publisher = {Oxford University Press},
  address   = {New Delhi}
}

@article{cath-2018-governing,
  title={Governing artificial intelligence: ethical, legal and technical opportunities and challenges},
  author={Cath, Corinne},
  journal={Philosophical Transactions of the Royal Society A: Mathematical, Physical and Engineering Sciences},
  volume={376},
  number={2133},
  year={2018},
  publisher={The Royal Society}
}

@book{mazzucato-2013-entrepreneurial,
  author    = {Mazzucato, Mariana},
  title     = {The Entrepreneurial State: Debunking Public vs. Private Sector Myths},
  year      = {2013},
  publisher = {Anthem Press},
  address   = {New York},
}

@article{jasanoff-2018-global,
  title={A global observatory for gene editing},
  author={Jasanoff, Sheila and Hurlbut, J Benjamin},
  journal={Nature},
  volume={555},
  number={7697},
  pages={435--437},
  year={2018},
  publisher={Nature Publishing Group UK London}
}

@article{winner-1980-artifacts,
 ISSN = {00115266},
 author = {Langdon Winner},
 journal = {Daedalus},
 number = {1},
 pages = {121--136},
 publisher = {The MIT Press},
 title = {Do Artifacts Have Politics?},
 urldate = {2026-02-26},
 volume = {109},
 year = {1980}
}

@book{ndlovu-gatsheni-2018-epistemic,
  author    = {Ndlovu-Gatsheni, Sabelo J.},
  title     = {Epistemic Freedom in Africa: Deprovincialization and Decolonization},
  year      = {2018},
  publisher = {Routledge},
  address   = {London}
}

@article{birhane-2021-algorithmic,
  author  = {Birhane, Abeba},
  title   = {Algorithmic Injustice: A Relational Ethics Approach},
  journal = {Patterns},
  year    = {2021},
  volume  = {2},
  number  = {2},
  pages   = {100205},
  doi     = {10.1016/j.patter.2021.100205}
}

@incollection{couldry-2019-costs,
  title={The costs of connection: How data is colonizing human life and appropriating it for capitalism},
  author={Couldry, Nick and Mejias, Ulises A},
  booktitle={The costs of connection},
  year={2019},
  publisher={Stanford University Press}
}

@article{jasanoff-2003-technologies,
  title={Technologies of humility: Citizen participation in governing science},
  author={Jasanoff, Sheila},
  journal={Minerva},
  volume={41},
  number={3},
  pages={223--244},
  year={2003},
  publisher={Springer}
}

@book{nye-1996-american,
  title={American technological sublime},
  author={Nye, David E},
  year={1996},
  publisher={mit Press}
}

@book{hughes-1993-networks,
  title={Networks of power: electrification in Western society, 1880-1930},
  author={Hughes, Thomas Parke},
  year={1993},
  publisher={JHU press}
}

@book{noble-1979-america,
  title={America by design: Science, technology, and the rise of corporate capitalism},
  author={Noble, David F},
  number={588},
  year={1979},
  publisher={Oxford University Press}
}

@book{jasanoff-1998-fifth,
  title={The fifth branch: science advisers as policymakers},
  author={Jasanoff, Sheila},
  year={1998},
  publisher={Harvard University Press}
}

@book{streeter-2010-net,
  title={The net effect: Romanticism, capitalism, and the Internet},
  author={Streeter, Thomas},
  volume={32},
  year={2010},
  publisher={NYU Press}
}

@book{abbate-2000-inventing,
  title={Inventing the internet},
  author={Abbate, Janet},
  year={2000},
  publisher={MIT press}
}
\end{document}